\documentclass[aps,prl,reprint,superscriptaddress,nofootinbib,nobibnotes]{revtex4-2}
\usepackage{amsmath}
\usepackage{graphicx}
\usepackage{multirow}
\usepackage[colorlinks=true,linkcolor=blue,urlcolor=blue,citecolor=blue]{hyperref}
\usepackage{makecell}
\usepackage{listings} 

\begin{document}
\title{Any Light Particle Searches with ALPS\,II: first science results}

\author{Daniel C. Brotherton}
\affiliation{Department of Physics, University of Florida, 32611 Gainesville, Florida, USA}

\author{Zachary R. Bush}
\altaffiliation{Now at: https://www.ionq.com/}
\affiliation{Department of Physics, University of Florida, 32611 Gainesville, Florida, USA}

\author{Sandy Croatto}
\affiliation{Deutsches Elektronen-Synchrotron DESY, Notkestr. 85, 22607 Hamburg, Germany}

\author{Mauricio Diaz-Ortiz Jr.}
\altaffiliation{Now at: 
Donders Institute - Biophysics, Radboud University, 6525 AJ, Nijmegen, Nederland}
\affiliation{Department of Physics, University of Florida, 32611 Gainesville, Florida, USA}

\author{Jacob Egge}
\affiliation{Deutsches Elektronen-Synchrotron DESY, Notkestr. 85, 22607 Hamburg, Germany}

\author{Aldo Ejlli}
\affiliation{Max-Planck-Institut für Gravitationsphysik (Albert-Einstein-Institut) and Leibniz Universität Hannover, 30167 Hannover, Germany}

\author{Henry Frädrich}
\affiliation{Deutsches Elektronen-Synchrotron DESY, Notkestr. 85, 22607 Hamburg, Germany}

\author{Joe Gleason}
\affiliation{Department of Physics, University of Florida, 32611 Gainesville, Florida, USA}

\author{Hartmut Grote}
\affiliation{School of Physics and Astronomy, Cardiff University, Cardiff CF24 3AA, United
Kingdom}

\author{Ayman Hallal}
\altaffiliation{Now at: https://optiwave.com/}
\affiliation{Max-Planck-Institut für Gravitationsphysik (Albert-Einstein-Institut) and Leibniz Universität Hannover, 30167 Hannover, Germany}

\author{Michael T. Hartman}
\altaffiliation{Now at: Institute Fresnel, 13397 Marseille, France}
\affiliation{Deutsches Elektronen-Synchrotron DESY, Notkestr. 85, 22607 Hamburg, Germany}

\author{Harold Hollis}
\affiliation{Department of Physics, University of Florida, 32611 Gainesville, Florida, USA}

\author{Katharina-Sophie Isleif}
\affiliation{Helmut-Schmidt-Universität, 22043 Hamburg, Germany}

\author{Alasdair L. James}
\altaffiliation{Now at: LIGO Laboratory, California Institute of Technology, Pasadena, CA 91125, USA}
\affiliation{School of Physics and Astronomy, Cardiff University, Cardiff CF24 3AA, United
Kingdom}

\author{Friederike Januschek}
\affiliation{Deutsches Elektronen-Synchrotron DESY, Notkestr. 85, 22607 Hamburg, Germany}

\author{Kanioar Karan}
\affiliation{Max-Planck-Institut für Gravitationsphysik (Albert-Einstein-Institut) and Leibniz Universität Hannover, 30167 Hannover, Germany}

\author{Sven Karstensen}
\author{Todd Kozlowski}
\author{Axel Lindner}
\email{axel.lindner@desy.de}
\affiliation{Deutsches Elektronen-Synchrotron DESY, Notkestr. 85, 22607 Hamburg, Germany}

\author{Giuseppe Messineo}
\altaffiliation{Now at: Sezione di Padova, Istituto Nazionale di Fisica Nucleare, 35131 Padua, Italy}
\affiliation{Department of Physics, University of Florida, 32611 Gainesville, Florida, USA}

\author{Manuel Meyer}
\affiliation{CP3-origins, Department of Physics, Chemistry and Pharmacy, University of Southern Denmark, Campusvej 55, 5230 Odense, Denmark}

\author{Guido Müller}
\affiliation{Department of Physics, University of Florida, 32611 Gainesville, Florida, USA}
\affiliation{Max-Planck-Institut für Gravitationsphysik (Albert-Einstein-Institut) and Leibniz Universität Hannover, 30167 Hannover, Germany}

\author{Ryan Netrval}
\affiliation{Max-Planck-Institut für Gravitationsphysik (Albert-Einstein-Institut) and Leibniz Universität Hannover, 30167 Hannover, Germany}

\author{Isabella Oceano}
\altaffiliation{Now at: Universität Hamburg, 22761 Hamburg, Germany}
\affiliation{Deutsches Elektronen-Synchrotron DESY, Notkestr. 85, 22607 Hamburg, Germany}

\author{Gulden Othman}
\altaffiliation{Now at: https://munich-quantum-instruments.com/}
\affiliation{Helmut-Schmidt-Universität, 22043 Hamburg, Germany}

\author{Jan H. P\~old}
\altaffiliation{Now at: https://www.leuze.com}
\affiliation{Deutsches Elektronen-Synchrotron DESY, Notkestr. 85, 22607 Hamburg, Germany}

\author{David Reuther}
\author{Andreas Ringwald}
\affiliation{Deutsches Elektronen-Synchrotron DESY, Notkestr. 85, 22607 Hamburg, Germany}

\author{Elmeri Rivasto}
\affiliation{CP3-origins, Department of Physics, Chemistry and Pharmacy, University of Southern Denmark, Campusvej 55, 5230 Odense, Denmark}

\author{Jos\'e Alejandro Rubiera Gimeno}
\affiliation{Helmut-Schmidt-Universität, 22043 Hamburg, Germany}

\author{Jörn Schaffran}
\author{Uwe Schneekloth}
\altaffiliation{Now at: Physikalisches Institut der Universit\"at Bonn, 53115 Bonn, Germany}
\author{Christina Schwemmbauer}
\altaffiliation{Now at: https://munich-quantum-instruments.com/}
\author{Richard C.G. Smith}
\altaffiliation{Now at: https://aerospace.org/}
\author{Aaron D. Spector}
\affiliation{Deutsches Elektronen-Synchrotron DESY, Notkestr. 85, 22607 Hamburg, Germany}

\author{David B. Tanner}
\affiliation{Department of Physics, University of Florida, 32611 Gainesville, Florida, USA}

\author{Dieter Trines$\,^\mathrm{a}$}
\affiliation{Deutsches Elektronen-Synchrotron DESY, Notkestr. 85, 22607 Hamburg, Germany}
\footnote{$\,^\mathrm{a\,}$deceased in July 2023}

\author{Li-Wei Wei}
\altaffiliation{Now at: MPI für Gravitationsphysik (AEI) and Leibniz Universität Hannover, 30167 Hannover, Germany}
\affiliation{Deutsches Elektronen-Synchrotron DESY, Notkestr. 85, 22607 Hamburg, Germany}

\author{Benno Willke}
\affiliation{Max-Planck-Institut für Gravitationsphysik (Albert-Einstein-Institut) and Leibniz Universität Hannover, 30167 Hannover, Germany}

\author{Rachel Wolf}
\affiliation{Deutsches Elektronen-Synchrotron DESY, Notkestr. 85, 22607 Hamburg, Germany}

\collaboration{ALPS\,II collaboration}

\begin{abstract}
The light-shining-through-a-wall experiment ALPS\,II at DESY in Hamburg searched for axions and similar lightweight particles in its first science campaign from February to May 2024. 
No evidence for the existence of such particles was found. For pseudoscalar bosons like the axion, with masses below about 0.1\,meV, we achieved a limit of $\mathrm{}{|g_{\phi\gamma\gamma}^p| < 1.5\cdot 10^{-9}\,\mathrm{GeV^{-1}} }$ at a 95\% confidence level for the di-photon coupling strength. 
This is more than a factor of 20 improvement compared to all previous similar experiments. 
We also provide limits on photon interactions for scalar, vector and tensor bosons. 
An achievement of this first science campaign is the demonstration of stable operation and robust calibration of the complex experiment. 
Currently, the optical system of ALPS\,II is being upgraded aiming for another two orders of magnitude sensitivity increase. 
\end{abstract}

\maketitle

\section{Introduction}
In spite of overwhelming evidence for the necessity of physics beyond the precisely tested Standard Model of particle physics and the technical successes of projects probing the high energy frontier directly, we lack any experimental evidence for an additional energy scale between the electroweak and the Planck scale (see summaries in \cite{ParticleDataGroup:2024cfk}). 
Hence, interest in accessing higher energy scales indirectly, for example via the search for pseudo-Goldstone bosons, is rising quickly.
The most famous example of such a boson is the axion \cite{Peccei:1977hh,Weinberg:1977ma,Wilczek:1977pj}, primarily motivated by an explanation of CP conservation in QCD. 
From astrophysical observations and particle physics experiments, it was quickly noticed
that the breaking of the global symmetry giving rise to the axion must occur at very high energies $\mathrm{}{f_{a}}$ above the electroweak scale (see \cite{Sikivie:1983ip} and references therein).
This axion was named ``invisible" as its coupling strengths to Standard Model constituents are predicted to be proportional to $\mathrm{}{f_{a}^{-1}}$.
While $\mathrm{}{f_{a}}$ being much larger than the electroweak scale comes with huge experimental challenges, it makes the axion an ideal cold dark matter candidate
\cite{Abbott:1982af,Preskill:1982cy,Dine:1982ah,Arias:2012az,DiLuzio:2020wdo,Chadha-Day:2021szb,Eroncel:2022vjg},
adding strong cosmological motivation to the particle physics science case.
Furthermore, axion-like particles have been proposed to explain a small cosmological constant and generate a naturally small electroweak scale \cite{Graham:2019bfu,Espinosa:2015eda}.
Recent results from the DESI large-scale cosmological survey have again sparked discussions on axion dark energy (see for example \cite{Seto:2024cgo,Qu:2024lpx}). 

In general, many theories for physics beyond the Standard Model predict the existence of axions, axion-like particles, hidden photons, or other WISPs (Weakly Interacting Slim Particles, introduced by A.\,Ringwald in 2007 at the 3rd Joint ILIAS–CERN–DESY Axion–WIMPs workshop) 
\cite{Arvanitaki:2009fg,Cicoli:2012sz,Jaeckel:2010ni,Hui:2021tkt,Dvali:2024zpc}.
Due to their extremely weak interactions, such WISPs cannot be found at existing or planned accelerator-based experiments. 
Instead, dedicated setups are needed, which are reviewed in \cite{Sikivie:2020zpn}. 
World-wide, three approaches are followed:
\begin{itemize}
	\item Haloscopes target WISPs as a component of our local dark matter.
	\item Helioscopes search for WISPs emitted by the Sun.
	\item Purely laboratory-based experiments try to generate and detect WISPs within one setup.
\end{itemize} 
The Any-Light-Particle-Search II (ALPS\,II) experiment at DESY in Hamburg is a prime example for the third category, one which is independent of astrophysical and cosmological assumptions. 
Within its particle physics strategy, DESY also plans to host the helioscope Baby\-IAXO~\cite{IAXO:2020wwp} and the haloscope MADMAX~\cite{MADMAX:2024jnp,MADMAX:2024sxs}. 

ALPS\,II follows the light-shining-through-a-wall (LSW) approach \cite{Redondo:2010dp,Arias:2010bh}:
\begin{itemize}
	\item Light converts to axions when passing a magnetic dipole field.
	\item The light is blocked by a light-tight wall, which will be passed unhindered by any axion.
	\item In a magnetic dipole field behind the wall, some of the axions convert back to photons.
\end{itemize} 
ALPS\,II builds upon experiences with ALPS \cite{ALPS:2009des,Ehret:2010mh}, the first LSW experiment at DESY.
At design sensitivity, it will surpass the axion-photon coupling-strength sensitivity of OSQAR at CERN \cite{OSQAR:2015qdv}, the previously most sensitive LSW-experiment, by three orders of magnitude.

\section{Boson detection in LSW experiments}
LSW experiments provide probabilities (or upper limits) for photons to seemingly pass light-tight walls. 
In the following, we briefly outline how to determine properties of scalar and pseudoscalar, vector and tensor bosons from such measurements. Using ALPS\,II data to look for minicharged particles like in \cite{Ehret:2010mh} or high-frequency gravitational waves (compare \cite{Ejlli:2019bqj,Ringwald:2020ist}) as well as more complex scenarios involving several axion-like particles \citep{deGiorgi:2025ldc} will be addressed in future publications.

For the probabilities of photon-boson conversions, 
\begin{equation}{
\mathcal{P}_{\gamma \rightarrow \mathrm{WISP} }
= \mathcal{P}_{\mathrm{WISP} \rightarrow \gamma}
=: \mathcal{P}_{\gamma \leftrightarrow \mathrm{WISP}}
}\end{equation}\
holds true for all interaction Lagrangians below. 

\subsection{Scalar and pseudoscalar bosons}
We follow references \cite{Redondo:2010dp,Arias:2010bh}, where the interaction of scalar and pseudoscalar bosons with electromagnetic fields is described by effective Lagrangians including 
\begin{align}
\mathcal{L} &\supset \frac{1}{4}g_{\phi \gamma \gamma} \cdot \phi \cdot F_{\mu\nu} F^{\mu\nu}
\ \mathrm{(scalars)}, \\
\mathcal{L} &\supset \frac{1}{4}g_{\phi \gamma \gamma} \cdot \phi \cdot F_{\mu\nu} \tilde{F}^{\mu\nu}
\ \mathrm{(pseudoscalars).}
\end{align}
with the photon-boson coupling strength $\mathrm{}{g_{\phi \gamma \gamma}}$, the boson field $\mathrm{}{\phi}$ and the electromagnetic field tensor $\mathrm{}{F_{\mu\nu}}$ as well as its dual 
$\mathrm{}{\tilde{F}_{\mu\nu} = \frac{1}{2}\epsilon_{\mu\nu\rho\lambda}F^{\rho\lambda}}$.
The probability for $\mathrm{}{\phi}-\gamma$ conversion in vacuum then reads:
\begin{subequations}\label{eq:scalar}
\begin{equation}\tag{\ref{eq:scalar}}
\mathcal{P}_{\gamma \leftrightarrow \phi} 
= \frac{1}{4} \frac{\omega}{k_\phi} 
\left( g_{\phi\gamma\gamma}BL \right)^2
\mid F_{N,\Delta}(qL) \mid ^2 \mathrm{\ with}
\end{equation}
\begin{equation}\label{eq:scalar-q}
q = \omega - \sqrt{\omega^2 - m_\phi^2},
\end{equation}
\vspace*{-0.5cm}
\begin{equation}\label{eq:scalar-F}
\mathrm{}{
\mid F_{N,\Delta}(qL) \mid =
\frac{2}{qL} \sin \left( \frac{qL}{2N} \right)
\frac{
\sin \left( \frac{qN}{2} \left( \frac{L}{N} + \Delta \right) \right)} 
{\sin \left( \frac{q}{2} \left( \frac{L}{N} + \Delta \right) \right)},
}\end{equation}
\end{subequations}
\begin{tabular}{lcl}
$\mathrm{}{\omega}$ & : & photon energy, \\
$\mathrm{}{k_\phi}$ & : & boson momentum, \\
$B,\, L$ & : & magnetic dipole field strength and length, \\
$\mathrm{}{m_\phi}$ & : & boson mass, \\
$N$, $\mathrm{}{\Delta}$ & : & number of dipoles, gap length between magnets.\\
\end{tabular}
\ \\
For scalar bosons, the photon polarization is perpendicular to the magnetic field direction
($\mathrm{}{\gamma_\perp}$), while pseudoscalar bosons, such as axions, require a parallel orientation
($\mathrm{}{\gamma_\parallel}$).

\subsection{Vector bosons}
Massive vector bosons $\mathrm{}{\gamma^\prime}$ (frequently named ``hidden'' or ``dark photons'') do not require any magnetic field for mixing with photons.
Following again \cite{Arias:2010bh}, one derives, with the kinetic mixing parameter $\mathrm{}{\epsilon}$ and the hidden vector potential $\mathrm{}{X_\mu}$, from 
\begin{equation}\mathrm{}{
\mathcal{L} \supset \frac{1}{2} \epsilon \cdot  X_{\mu\nu} F^{\mu\nu}
}
\end{equation}
the mixing probability in vacuum:
\begin{subequations}\label{eq:vector}
\begin{equation} \tag{\ref{eq:vector}}
\mathrm{}{
\mathcal{P}_{\gamma \leftrightarrow \gamma\prime}
= 4\, \epsilon^2 F_{\gamma^\prime}(qL_\mathrm{osci})},
\end{equation}
\vspace*{-0.5cm}
\begin{equation} \label{eq:vector-F}
F_{\gamma^\prime}(qL_\mathrm{osci}) = \sin^2 \left( \frac{qL_\mathrm{osci}}{2}  \right),
\end{equation}
\end{subequations}
with the mixing region length $\mathrm{}{L_\mathrm{osci}}$ and $q$ as in eq.\,\eqref{eq:scalar-q}.

\subsection{Tensor bosons}
\label{subsec:tensors}
For the interaction of photons with massive parity-even spin-2 bosons
$\chi$ in a magnetic field, we follow the formalism of \cite{Garcia-Cely:2025ula}:
\begin{equation}{
\mathcal{L} \supset \sqrt{ 8 \pi G^\prime} \cdot
\delta M_{\mu\nu}F^{\mu\alpha} F^\nu_\alpha
}.\end{equation}
$G^\prime$ is the coupling strength and $\delta M_{\mu\nu}$ the tensor field.
The mixing probabilities depend on the polarization of light with respect to the magnetic field orientation. For a tensor boson mass $\mathrm{}{m \to 0}$, but $m > 0$, one finds:
\begin{subequations}
\begin{align}
\mathcal{P}_{\gamma_\perp \leftrightarrow \chi}
&= \frac{4}{3} \cdot  4 \pi G^\prime
\left( BL \right) ^2
\mid F_{N,\Delta}(qL) \mid ^2,
\label{eq:tensor-perp}\\
\mathcal{P}_{\gamma_\parallel \leftrightarrow \chi}
&= 4 \pi G^\prime
\left(BL \right)^2
\mid F_{N,\Delta}(qL) \mid ^2, \ \ \ 
\label{eq:tensor-par}
\end{align}
\end{subequations}
where q and $\mathrm{}{F_{N,\Delta}(qL)}$ are defined as in eq.\,\eqref{eq:scalar-q} and \eqref{eq:scalar-F}.

\subsection{Standard model background}
An LSW effect can be mediated by photon-graviton conversion in a background magnetic field. With Newton's constant $G$, this probability is estimated as (see \cite{Raffelt:1987im} and \cite{Krizova:2024lvy}, eq.\,3.157): 
\begin{equation}
\mathcal{P}_\mathrm{LSW}
\simeq \left( 4\pi G \right)^2 \left( B L \right)^4 \\
=  7\cdot 10^{-75} \left( B[\mathrm{T}]\cdot L[\mathrm{m}] \right)^4.
\end{equation}
At the  ALPS\,II design sensitivity \cite{Bahre:2013ywa}, 
this results in a power of 
$\mathrm{}{\approx 4\cdot 10^{-54}}$\,W behind the wall, which is totally negligible for any practical purposes. 

\section{The ALPS\,II experiment}
The experiment ALPS\,II is located in a straight section of the former HERA accelerator complex \cite{19436} around the hall North at DESY in Hamburg.
ALPS\,II, as laid out in \cite{Bahre:2013ywa}, is striving for the first realization of an optical resonantly enhanced LSW experiment proposed earlier \cite{Hoogeveen:1990vq,Fukuda:1996kwa,Sikivie:2007qm}. 
It consists of two strings of modified superconducting dipole magnets (originally built for the HERA proton beam) before and after the wall as well as three cleanrooms at both ends and in the center to host the optical components. 

\subsection{The magnet strings}
The magnets and the infrastructure of HERA, shut down in 2007, provide a unique environment for a large LSW experimental setup, which was first noticed in \cite{Ringwald:2003nsa}.
ALPS\,II relies especially on the following opportunities:
\begin{itemize}
	\item About 300\,m long straight sections in the HERA tunnel allow for the installation of 24 superconducting HERA dipole magnets.
	\item Infrastructure and the cryogenics for cooling the magnets \cite{Horlitz:1984ym} could be re-used with limited modifications. 
	\item The cold bore of the HERA dipoles, with a diameter of 55\,mm, is, in principle, large enough to allow for a several hundred meter long high-finesse optical resonator. 
\end{itemize}
However, the cold mass inside an unmodified HERA dipole magnet is bent as the magnet was constructed to guide protons through the arcs of the accelerator. This would have limited the horizontal free aperture to 35\,mm, insufficient for a high-finesse optical cavity making full use of a HERA straight section. Therefore, a ``brute-force'' straightening procedure was developed \cite{Albrecht:2020ntd}. Without removing the cold mass from the cryostat, suspensions of the cold mass were changed and special pressure-props installed at the positions of cryostat flanges to recover, on average, a 50\,mm horizontal aperture. It is remarkable that the straightened dipoles show essentially the same performance specification in tests as at the time of their fabrication about 40 years ago \cite{Albrecht:2020ntd}.
This success allowed us to install 12 plus 12 straightened magnets in the straight tunnel section around the HERA North hall.
The parameters of the magnet strings are summarized in Table\,\ref{tab:Dipoleparams}. 
Note that the magnetic lengths in front of and behind the wall are equal: 
$\mathrm{}{L = N \cdot L_\mathrm{SD}}$.
\begin{table}[!h]
	\centering
	\begin{tabular}{|l|r|}
	\hline
	Parameter & Value \\ \hline
	Single dipole magnetic length & $\mathrm{}{L_\mathrm{SD}=(8826\pm 2)\,\mathrm{mm}}$ \\
	Dipole magnetic field strength & $\mathrm{}{B=(5.318\pm0.005)\,\mathrm{T}}$ \\
	Number of dipoles & $\mathrm{}{2\cdot N = 2\cdot12}$ \\
	Gap between dipoles & $\mathrm{}{\Delta=(936\pm2)\,\mathrm{mm}}$ \\
	Gap between strings & (6290$\pm$5)\,mm \\
	\hline
	\end{tabular}
	\caption{Parameters of the magnet strings.}
	\label{tab:Dipoleparams}
\end{table}

\subsection{The optical system}
\begin{figure*}[tp]
    \centering
    \includegraphics[width=1.0\linewidth]{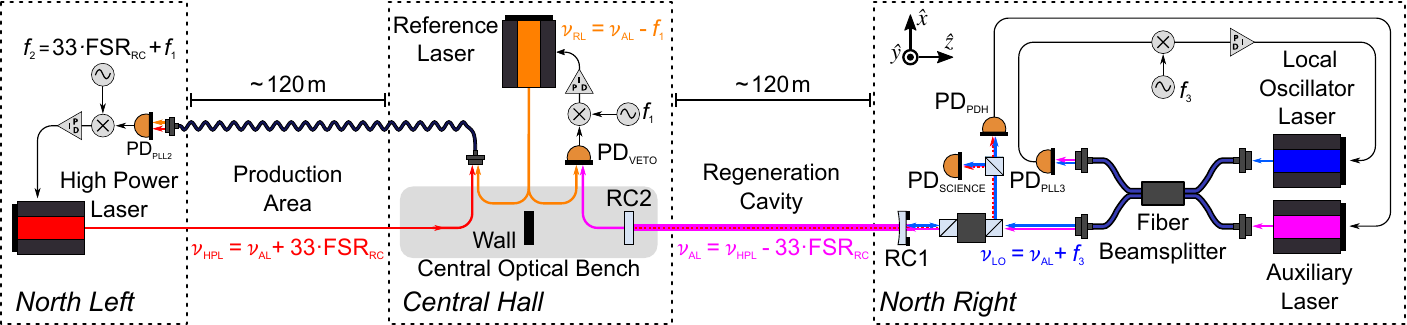}
    \caption{The optical system implemented in the first science run of ALPS\,II (taken from \cite{Spector:2026eys}). ``FSR" denotes the free spectral range of the regeneration cavity, ``North Left", ``Central Hall" and ``North Right" refer to the three cleanrooms at the ALPS\,II magnet strings. The black bar on the ``central optical bench" indicates the light-tight wall housing a shutter which is opened for system checks and calibration. See the text for a brief explanation and \cite{Spector:2026eys} for details.
Twelve HERA dipole magnets each surround the production area and the regeneration cavity.}
    \label{fig:optics-system}
\end{figure*}

While the final optical system of ALPS\,II will feature two mode-matched optical resonators \citep{Ortiz:2020tgs}, we only implemented the regeneration cavity (RC) behind the wall \cite{Kozlowski:2024jzm} for the first science run. This was done to simplify the operation and to increase potential stray-light ``leak'' intensities through the wall by a factor of 40 so that they could be traced more easily. Fig.\,\ref{fig:optics-system} sketches the optics system. Details of the system, its calibration and data analyses will be presented in a forthcoming publication\,\cite{Spector:2026eys}.

The frequency of the auxiliary laser (purple in Fig.\,\ref{fig:optics-system}) is constantly adjusted to stay on resonance with the regeneration cavity, compensating for ambient fluctuations of the length of the regeneration cavity.  
The transmitted field is used to phase-lock the reference laser (orange in Fig.\,\ref{fig:optics-system}) to the auxiliary laser at a fixed frequency offset. 
The offset is selected such that light from the reference laser is not resonant in the regeneration cavity. 
The frequency of the high-power laser (red in in Fig.\,\ref{fig:optics-system}) is in turn phase-locked to the reference laser. The difference frequency is chosen to ensure that the laser is resonant in the regeneration cavity; offset by 33 FSRs from the auxiliary laser.
This also ensures that light regenerated from new bosons behind the wall is resonant within the cavity.
The optical layout was carefully designed to avoid any injection of light from the high power laser into the regeneration cavity when the shutter (aka the wall in Fig.\,\ref{fig:optics-system}) was closed. 

The frequency of the local oscillator laser (blue in Fig.\,\ref{fig:optics-system}) is again stabilized with respect to the auxiliary laser with a frequency offset such that it is not resonant with the regeneration cavity. 
The local oscillator laser light reflected from the cavity interferes with the light (potentially) regenerated after the wall that is leaving the cavity as it transmits the mirror RC1. 
A heterodyne sensing scheme \cite{Spector:2026eys,Hallal:2020ibe} is used to sense the regenerated power via a measurement of the interference beatnote between these two fields at the photodetector $\mathrm{PD_{science}}$.   
If a signal were to appear, the beatnote frequency  would be the known fixed frequency difference between the local oscillator laser and the high-power laser, referred to here as the heterodyne frequency $f_\mathrm{s}$.

The most relevant parameters of the optical system used here are given in Table\,\ref{tab:Opticsparams}. 
We note that the collaboration has developed an alternative optical scheme using a Transition Edge Sensor detector system \cite{Gimeno:2023nfr} to count single photons generated behind the wall. This system could be implemented at ALPS\,II to cross-check a discovery with the heterodyne sensing scheme described here.
\begin{table}[!h]
	\centering
	\begin{tabular}{|l|r|}
	\hline
	Parameter & Value \\ \hline
	Wavelength & ($1064.5\pm 0.1$)\,nm \\
	Vacuum-tube length bf.\ wall & $\mathrm{}{L_\mathrm{V} = (123.00\pm 0.01 )\,}$m \\
	Cavity length behind wall & $\mathrm{}{L_\mathrm{RC} = (122.6012 \pm 0.0001) \,}$m \\
	Laser power injected & $\mathrm{}{P_\mathrm{i} \approx 25\,}$W \\
	Cavity res.\,enhancement  & $\mathrm{}{\beta \approx 7000}$ \\
	Cavity free spectral range & $\mathrm{FSR} \approx 1.22263\,$MHz \\
	 Spatial \& spectral matching& 
	 \multirow{2}{*}{$|\eta|^2 \approx 0.9$}\\
	 of HPL to the RC Eigenmode & \\
	 Vacuum system gas pressure &$< 10^{-9}$\,mbar \\
	\hline
	\end{tabular}
	\caption{Parameters of the optical system. $\mathrm{}{P_\mathrm{i}}$ denotes the laser power traversing the magnet string before the wall.
$\mathrm{}{P_\mathrm{i}}$, $\beta$ and  $|\eta|^2 $ varied during data taking (see \cite{Spector:2026eys}). $L_\mathrm{RC}$ is measured via the free spectral range of the cavity.}
	\label{tab:Opticsparams}
\end{table}

\section{Data analysis}
A shutter in the light-tight wall is used to help calibrate ALPS\,II.
It can be opened to allow a tiny fraction of the HPL light to enter the regeneration cavity. 
Then, the probability of a photon-WISP conversion 
$ \mathrm{}{
\mathcal{P}_{\gamma \leftrightarrow \mathrm{WISP}}
}$
is experimentally determined by comparing the powers at the HPL frequency reaching the science photodetector $\mathrm{{PD_{science}}}$ (behind the cavity mirror $\mathrm{{{RC1}}}$ in Fig.\,\ref{fig:optics-system}) with the shutter in the wall open and closed. 
The powers are determined by evaluating the heterodyne beat signal.
The procedure is sketched below. Details can be found in \cite{Spector:2026eys}.
We use the following notations for light powers at the HPL frequency:\\
\begin{tabular}{lcl}
$\mathrm{}{P_\mathrm{i}}$ & : & Power injected by the high-power laser\\
$\mathrm{}{P_\mathrm{open}}$ & : & Power measured with open shutter \\
$\mathrm{}{P_\gamma}$ & : & Power measured with closed shutter\\
\end{tabular}

\subsection{Open-shutter mode}
The cavity finesse is determined by its two mirrors $\mathrm{}{\mathrm{RC2}}$ and $\mathrm{}{\mathrm{RC1}}$ with transmissivities $\mathrm{}{T_\mathrm{RC2}}$ and ${T_\mathrm{RC1}}$ and extra optical losses $\mathrm{}{l}$.
In open shutter mode, the HPL light enters the cavity via the mirror $\mathrm{}{\mathrm{RC2}}$.
The power injected into the cavity is given by $\mathrm{}{P_\mathrm{i}}$ and the transmissivity of the central optical bench $\mathrm{}{T_\mathrm{COB}}$ (the light has to pass four high reflective mirrors, not shown in Fig.\,\ref{fig:optics-system}) before reaching $\mathrm{{{RC2}}}$. 
The transmitted power behind $\mathrm{}{\mathrm{RC1}}$ is given by
\begin{equation}
P_\mathrm{open} = |\eta|^2 \cdot T_\mathrm{COB} \cdot T_\mathrm{RC} \cdot P_\mathrm{i}.
\end{equation}
Here $|\eta|^2$ denotes the spatial and spectral matching of the high power laser to the cavity Eigenmode, and
$T_\mathrm{RC}$ the transmissivity of the regeneration cavity. 
For a high-finesse cavity, $T_\mathrm{RC}$  is approximated as
\begin{equation}\mathrm{}{
T_\mathrm{RC} \simeq \frac{4\cdot T_\mathrm{RC1} \cdot T_\mathrm{RC2}}{(T_\mathrm{RC1}+T_\mathrm{RC2}+l)^2}}.
\end{equation}

\subsection{Closed-shutter mode}
The re-conversion of bosons to light is resonantly enhanced by the cavity \cite{Hoogeveen:1990vq,Fukuda:1996kwa,Sikivie:2007qm}.
This light generated inside the cavity leaves the cavity through one of its mirrors,
or otherwise through a cavity loss mechanism. 
The resonant enhancement factor for light generated inside the cavity and sensed behind $\mathrm{{{RC1}}}$ is approximately
\begin{equation}
\beta \simeq \frac{4\cdot T_\mathrm{RC1}}{(T_\mathrm{RC1}+T_\mathrm{RC2}+l)^2} 
\simeq \frac{T_\mathrm{RC}}{T_\mathrm{RC2}}.
\end{equation}
Hence, the power at the HPL frequency measured with closed shutter is
\begin{subequations}
\begin{align}
P_\gamma &= 
\mathcal{P}_{\gamma \leftrightarrow \mathrm{WISP}}^2  \cdot 
|\eta|^2 \cdot \beta \cdot P_\mathrm{i}
= \mathcal{P}_{\gamma \leftrightarrow \mathrm{WISP}}^2  \cdot
|\eta|^2 \cdot \frac{T_\mathrm{RC}}{T_\mathrm{RC2}} \cdot P_\mathrm{i} \notag \\
&=
\mathcal{P}_{\gamma \leftrightarrow \mathrm{WISP}}^2  \cdot
\frac{P_\mathrm{open}}{T_\mathrm{COB} T_\mathrm{RC2}} \tag{\theequation}
\end{align}
\end{subequations}
resulting in
\begin{equation}
\mathcal{P}_{\gamma \leftrightarrow \mathrm{WISP}} = 
\sqrt{\frac{P_{\gamma}}{P_\mathrm{open}} T_\mathrm{COB}  T_\mathrm{RC2}}
\end{equation}
Most systematic uncertainties are identical in $\mathrm{}{P_\gamma}$ and $\mathrm{}{P_\mathrm{open}}$ and hence cancel in the ratio. Again, details on corresponding cross-checks are given in \cite{Spector:2026eys}.

\subsection{Experimental results}
In the first science campaign, over 580,000\,s and 1,060,000\,s of closed-shutter valid data were collected for injected laser light polarized perpendicular and parallel to the magnetic dipole field, respectively. 

\begin{figure}[h]
    \centering
    \includegraphics[width=1.0\linewidth]{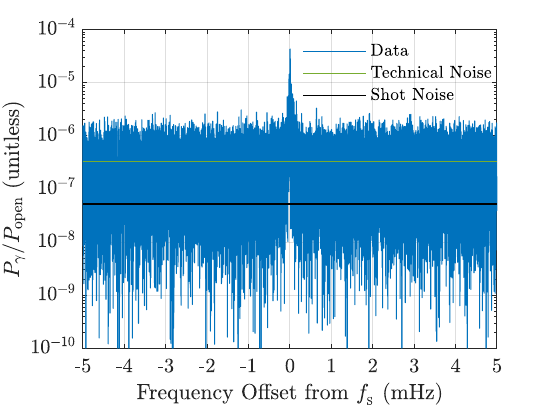}
    \caption{Taken from \cite{Spector:2026eys}: normalized spectral power from measurements on $\mathrm{{PD_{science}}}$ determined for frequencies around the signal's heterodyne frequency for the polarization $\mathrm{}{\gamma_\parallel}$ (Table\,\ref{tab:expresults}). An LSW signal would only show up in the frequency bin centered at 0\,Hz. The observed broader distribution is likely caused by stray-light.
    }
    \label{fig:pseudo-f}
\end{figure}

The main experimental results are exemplified in Fig.\,\ref{fig:pseudo-f}, which displays the spectral power normalized by the open shutter power, $\mathrm{}{P_{\gamma}(\Delta f_\mathrm{s}) /P_\mathrm{open}}$,
for frequency offsets $\Delta f_\mathrm{s}$ from the signals's heterodyne frequency $f_\mathrm{s}$. 
These normalized spectral power distributions were determined by evaluating the power modulation on $\mathrm{{PD_{science}}}$ at frequencies in the vicinity of $f_\mathrm{s}$.
A clear signal above background shows up at $\Delta f_\mathrm{s} = 0\,\mathrm{Hz}$. However, this peak is much broader than expected for LSW mediated by new bosons.
It is likely caused by stray-light passing around the wall (see \cite{Spector:2026eys}) with an intensity of a few $\mathrm{}{10^{-22}}$\,W at the HPL frequency.  
In Ref.\,\cite{Spector:2026eys}, 
the background at the signal frequency is estimated using the measured background at nearby frequencies.
The ratios of the results at the signal frequency to the estimated backgrounds are shown in column 3 of Table~\ref{tab:expresults}.
The results agree well with the expectation value of 1 in the absence of a signal.
In column 4 of the same table, we derive 95\% confidence level (CL) upper limits for the ratio $P_\gamma\, / \, P_\mathrm{open}$.
In Ref.\,\cite{Spector:2026eys}, it is shown that, at frequencies other than the signal frequency, the results adhere to a non-central $\mathrm{\chi^2}$ distribution.
The limits are therefore based on the assumption that the results at the signal frequency show a similar behavior.
Here, we define a 95\% CL as the probability to have not missed a detection corresponding to a significance of five standard deviations.\label{text:95limit}
Using the measurement of $\mathrm{}{T_\mathrm{COB}  T_\mathrm{RC2}}=(9.7\pm 1.2)\cdot 10^{-23}$, limits on the conversion probabilities $\mathrm{{\gamma \rightarrow WISP \rightarrow \gamma}}$ can be determined. 
Following \cite{COUSINS1992331}, we take a conservative approach and add the systematic uncertainties to the limits derived from statistical uncertainties, resulting in the limits on the conversion probabilities in column 4.\label{text:syslim}

\begin{table*}[tp]
	\centering
	\begin{tabular}{|c|c|c|c|c|}
	\hline
	Polarization & Duration & signal-bin\,/\,bckg.-estim.& 95\% CL UL $\left( P_\gamma\, / \, P_\mathrm{open}\right)$ & \rule{0pt}{10pt}95\% CL UL $\mathrm{}{\sqrt{\mathcal{P}_{\gamma \rightarrow \mathrm{WISP} \rightarrow \gamma}}}$  \\ \hline
	\rule{0pt}{9pt}$\mathrm{}{\gamma_\perp}$ & 580,000\,s & $\mathrm{}{0.85 \pm 1.03_\mathrm{\,stat}}$ & $\mathrm{}{(5.6 \pm 0.8_\mathrm{\,sys})\cdot 10^{-4}}$ & $\mathrm{}{2.5\cdot 10^{-13}}$  \\
	$\mathrm{}{\gamma_\parallel}$ & 1,060,000\,s & $\mathrm{}{2.67\pm 1.00_\mathrm{\,stat}}$ & ${(2.6 \pm 0.3_\mathrm{\,sys})\cdot 10^{-4}}$ & $\mathrm{}{1.7\cdot 10^{-13}}$  \\
	\hline
	\end{tabular}
	\caption{The main experimental results of the ALPS\,II first science campaign. ``Polarization'' is defined with respect to the orientation of the magnetic field. In the third column the power ratio of the closed- and open-shutter measurements at the HPL frequency is compared to a background estimation as explained in \cite{Spector:2026eys}. 
Here, the errors are statistical uncertainties as most of the systematic uncertainties cancel.
The fourth column displays upper limits on the power ratio of the closed- and open-shutter measurements with their systematic uncertainties.
The last column gives the upper limits on boson-photon conversion probabilities incorporating the systematic uncertainties (p.\,\pageref{text:syslim}).
Note also our definition of the 95\% CL on page\,\pageref{text:95limit}.}
	\label{tab:expresults}
\end{table*}

\section{WISP search results}
The results of the ALPS\,II first science campaign are summarized here and compared to likewise model-independent or nearly model-independent outcomes from other experiments. 
A full review of the quickly evolving WISPy experimental landscape is beyond the scope of this paper and we refer the reader to
reference \cite{AxionLimits} for an overview.
Table\,\ref{tab:ALPSII-couplings} summarizes the 95\% CL upper limits for $F_{N,\Delta}\rightarrow 1$ resp. $F_{\gamma^\prime} \rightarrow 1$ (equations \eqref{eq:scalar-F}, \eqref{eq:vector-F}).
\begin{table}[!h]
	\centering
	\begin{tabular}{|c|c|c|c|c|}
	\hline
	Boson & 95\%\,CL limit & see eq. & Figure  \\ \hline
	\rule{0pt}{9pt}pseudoscalar & $\mathrm{}{|g_{\phi\gamma\gamma}^p| < 1.5\cdot 10^{-9}\,\mathrm{GeV^{-1}} }$ & \multirow{2}{*}{\eqref{eq:scalar}} & \ref{fig:ALPSII-pseudo}\\
	scalar & $\mathrm{}{|g_{\phi\gamma\gamma}^s| < 1.8\cdot 10^{-9}\,\mathrm{GeV^{-1}} }$ & & \ref{fig:ALPSII-scalar} \\
	vector & $\mathrm{}{|\epsilon| < 2.0\cdot 10^{-7} }$ & \eqref{eq:vector} & \ref{fig:ALPSII-hp} \\
	tensor, $\mathrm{}{\gamma_\perp}$ & ${G^\prime / G < 7.3\cdot 10^{18}}$ & \eqref{eq:tensor-perp} &  \\
	tensor, $\mathrm{}{\gamma_\parallel}$ & ${G^\prime / G < 6.6\cdot 10^{18}}$ & \eqref{eq:tensor-par} &  \\
	tensor, combined & ${G^\prime / G < 5.8\cdot 10^{18}}$ & \eqref{eq:tensor-perp},\eqref{eq:tensor-par} & \ref{fig:ALPSII-tensor} \\
	\hline
	\end{tabular}
	\caption{95\% confidence level peak sensitivity limits for the couplings of different WISPs.
	The limits for tensor bosons coupling strengths $G^\prime$ is given relative to the standard-graviton coupling $G$.}
	\label{tab:ALPSII-couplings}
\end{table}

\begin{figure}[b]
    \centering
    \includegraphics[width=1.0\linewidth]{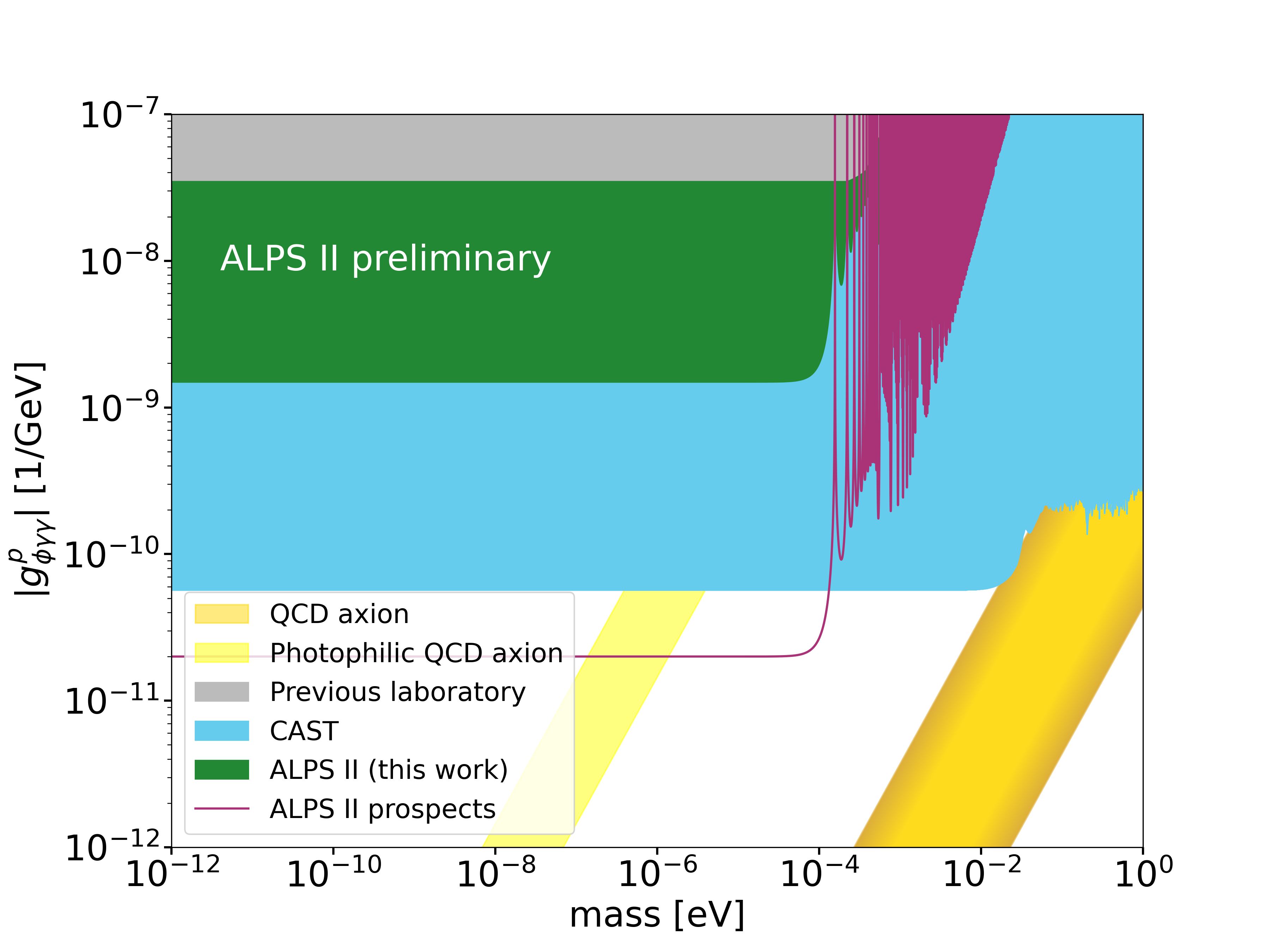}
    \caption{Limits on pseudoscalar bosons: ``Previous laboratory" (grey) summarizes previous results from ALPS\,\cite{Ehret:2010mh}, OSQAR\,\cite{OSQAR:2015qdv} and PVLAS\,\cite{Ejlli:2020yhk}; CAST (light-blue) is taken from \cite{CAST:2024eil} and references therein. The golden axion band shows an ``artist's" view on the approximate range given by KSVZ- and DFSZ-inspired models 
    \cite{PhysRevLett.43.103,SHIFMAN1980493,DINE1981199,Zhitnitsky:1980tq}, while the yellow range refers to a more recent model~\cite{Sokolov:2021ydn}. The green area shows the result of this analysis; the purple line, the ALPS\,II prospects.}
    \label{fig:ALPSII-pseudo}
\end{figure}

For pseudoscalar bosons (Fig.\,\ref{fig:ALPSII-pseudo}) the sensitivity achieved by ALPS\,II is about a factor of 20 better than at previous LSW experiments. 
ALPS\,II aims for surpassing the limits derived from solar axion searches (set by CAST) 
and will be sensitive to more recent QCD axion models involving e.g. magnetic monopoles~\cite{Sokolov:2021ydn}.

\begin{figure}[h]
    \centering
    \includegraphics[width=1.0\linewidth]{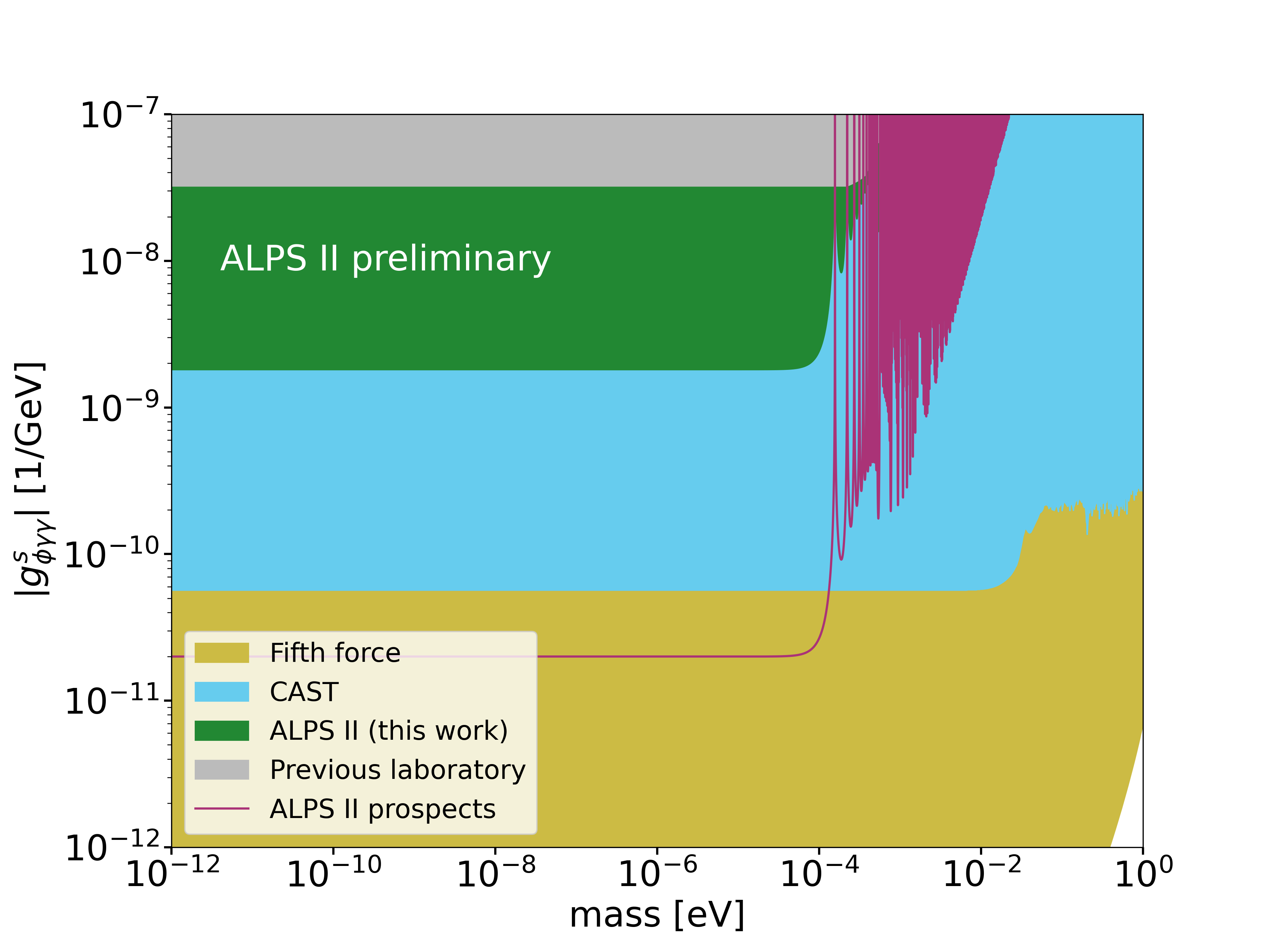}
    \caption{Limits on scalar bosons: ``Fifth force" (tan) is taken from \cite{Adelberger_2003,KONOPLIV2011401} with the scaling $\mathrm{}{Q_e \sim 1/500}$ used in  \cite{Arvanitaki_2016},
``Previous laboratory" (grey) summarizes results from ALPS\,\cite{Ehret:2010mh}, OSQAR\,\cite{OSQAR:2015qdv} and PVLAS\,\cite{Ejlli:2020yhk}; ``CAST'' (light-blue) is taken from \cite{CAST:2024eil} and references therein.
The green area shows the result of this analysis; the purple line, the ALPS\,II prospects.}
    \label{fig:ALPSII-scalar}
\end{figure}

For scalar bosons, we note that LSW experiments and similar very low mass WISP searches cannot compete with fifth force search sensitivities (Fig.\,\ref{fig:ALPSII-scalar}). 

\begin{figure}[!h]
    \centering
    \vspace{-0.5cm}
    \includegraphics[width=1.0\linewidth]{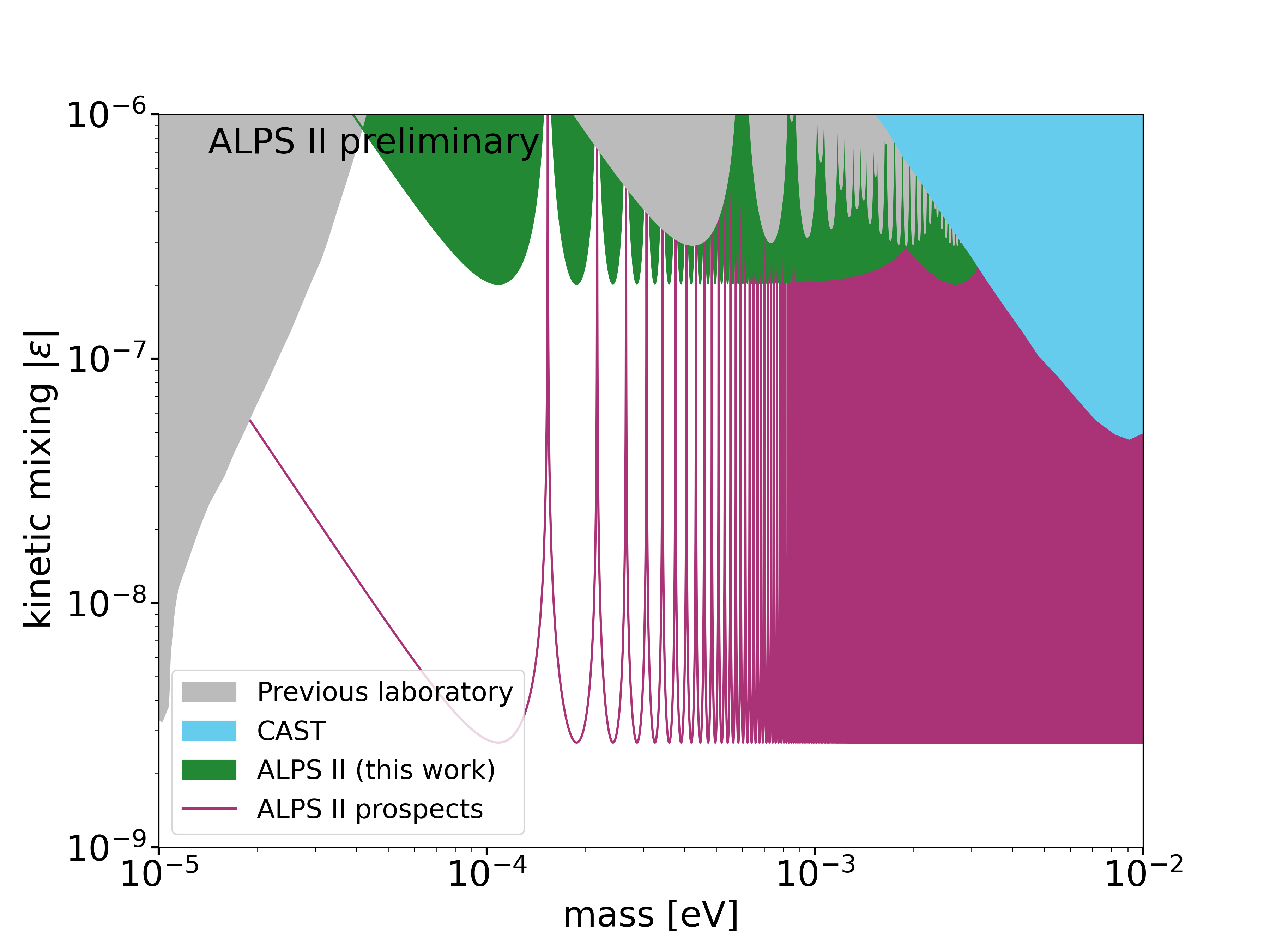}
    \caption{Limits on hidden photons: ``Previous laboratory" (grey) summarizes results from 
CROWS\,\cite{Betz_2013} and ALPS\,\cite{Ehret:2010mh}; ``CAST'' data (light-blue) are taken from \cite{Redondo_2008}.
The green area shows the result of this analysis; the purple line, the ALPS\,II prospects.}
    \label{fig:ALPSII-hp}
\end{figure}

The mixing of vector bosons with photons does not depend on the magnetic field.
Hence, the results of the two data sets with the different polarizations ($\gamma_\perp$ and $\gamma_\parallel$) can be combined.
We use a conservative approach by combining the limits on the power ratios of the closed- and open-shutter measurements
$P_\gamma\, / \, P_\mathrm{open}$ (without the systematic uncertainties) via
$\mathrm{{Limit(\perp + \parallel) = \left( Limit(\perp)^{-2}+Limit(\parallel)^{-2}\right) ^{-0.5}}}$
(see \cite{ParticleDataGroup:2024cfk}, section ``Statistics'').
The systematic uncertainty is then added to the combined result.
Note that in the first science campaign of ALPS\,II, 
${L_\mathrm{osci} = L_\mathrm{V}}$ before the wall and 
${L_\mathrm{osci} = L_\mathrm{RC}}$ behind it (eq.\,\eqref{eq:vector} and Tab.\,\ref{tab:Opticsparams}). 
As visible in Fig.\,\ref{fig:ALPSII-hp}, the improvement relative to earlier results is more modest, because the unique ALPS\,II magnet strings are irrelevant for vector boson searches.

\begin{figure}[htb]
    \centering
    \vspace{-0.5cm}
    \includegraphics[width=1.0\linewidth]{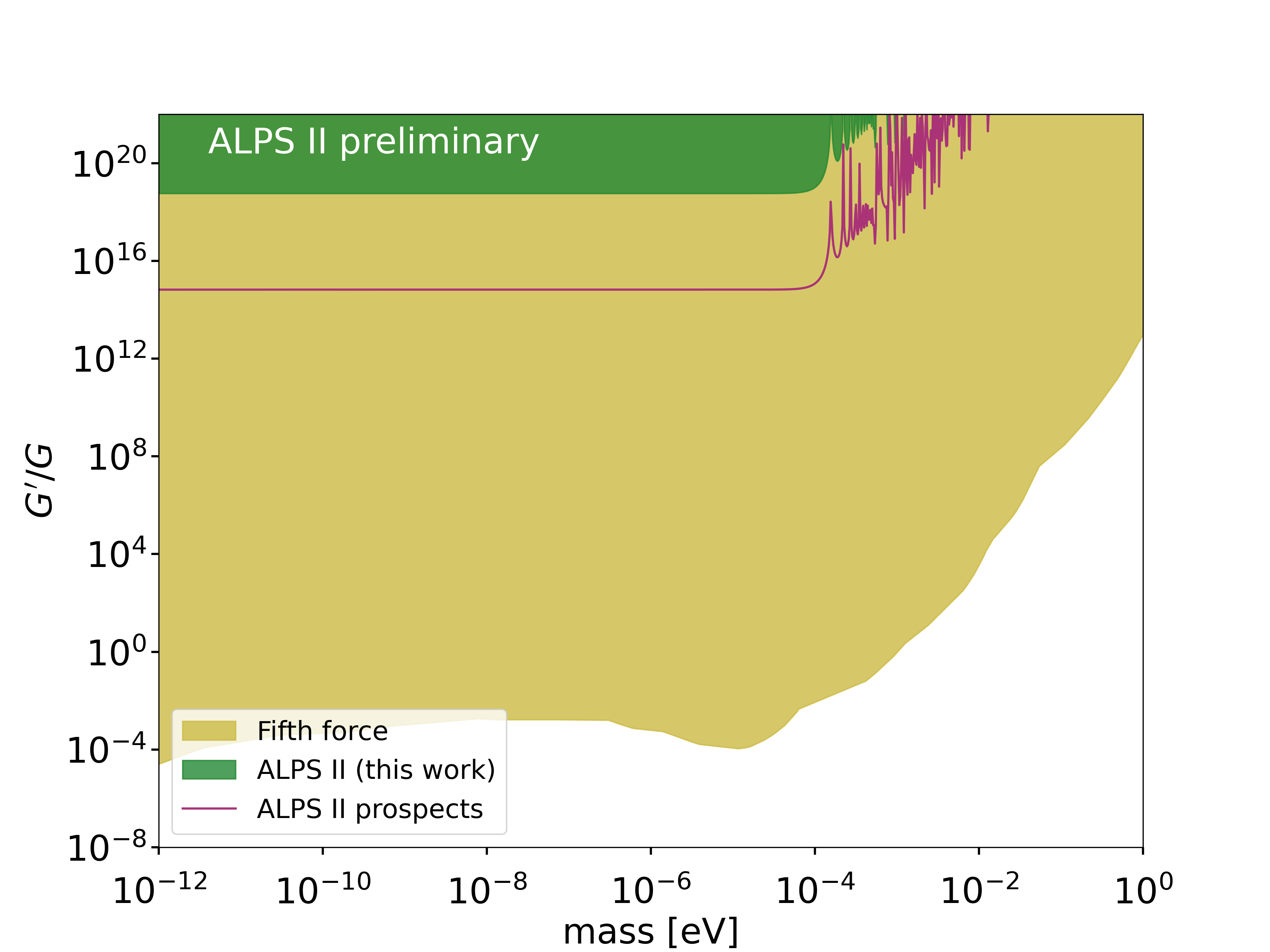}
    \caption{Limits on tensor bosons coupling strength $G^\prime$ relative to the standard-graviton coupling $G$.
    "Fifth force" (tan) is taken from \cite{Cembranos:2017vgi}.
    The green area shows the result of this analysis; the purple line, the ALPS\,II prospects.}
    \label{fig:ALPSII-tensor}
\end{figure}

Fig.\,\ref{fig:ALPSII-tensor} shows ALPS\,II exclusion limits on massive tensor bosons relative to the standard-graviton coupling strength $G = \mathrm{6.709\cdot 10^{-39}\,GeV^{-2}}$.
Here, we combined the limits from the two polarization states like described above, taking into account the different factors in equations \eqref{eq:tensor-perp} and \eqref{eq:tensor-par}.
Table \ref{tab:ALPSII-couplings} also lists the limits for the two tensor searches with different light polarizations. Note that in models involving Lorentz symmetry violation, tensor spin-0 polarization states do not propagate:
in this case, the limits from fifth force experiments in Fig.\,\ref{fig:ALPSII-tensor} do not apply (see \cite{zxtk-bwnf} for a discussion).

\section{Conclusions}
The ALPS\,II experiment, based  at DESY in Hamburg, Germany, is the first optical light-shining-through-a-wall experiment to exploit resonant regeneration. With the completion of the first science campaign it has successfully demonstrated stable data-taking and robust calibration.
Among other challenges, this required sub-\textmu Hz precision in the control of  the frequencies between the various lasers used in the setup 
and sensing of signals with powers below $\mathrm{10^{-22}\,W}$.

As a purely laboratory experiment, the ALPS\,II results do not depend on astrophysical or cosmological assumptions.
In the first data-taking campaign, no evidence for the existence of axions or others WISPs was found, although the sensitivity  on the axion-photon-coupling strength was increased by more than a factor of 20 compared to similar experiments. 

At present, the optics system of ALPS\,II is being upgraded with the goal to reach an axion-search sensitivity comparable to astrophysical analyses. 
The planned upgrades include mitigating the stray-light intensity on the science detector and its technical noise, implementing the production cavity, and, at a later stage, improving the finesse of the regeneration cavity. 

\section{Acknowledgments}
We are grateful to DESY for the unique opportunity to build ALPS\,II and thank especially the groups FE, MCS, MEA, MKS, MPC, MPS, MVS and ZM for their essential technical support. 
The ingenious procedure initiated by D.\,Trines to straighten the HERA dipole magnets provided the ALPS\,II basis. Unfortunately he was not able to witness the success of his work. Without the know-how, engagement and creativity of K.\,Gadow we would not have an LSW experiment in the HERA facilities.

We are thankful to numerous colleagues 
for their encouragement to realize ALPS\,II. Special thanks go to K.\,van\,Bibber for very valuable help in getting ALPS\,II off the ground and to C.\,O'Hare for his compilations of experimental results.
We acknowledge the support of the National Science Foundation (Grant No.\,1802006), of
the Heising-Simons Foundation (Grant No.\,2015-154 and 2020-1841), of the Deutsche Forschungsgemeinschaft through project grant WI 1643/2-1 and EXC 2121 ``Quantum
Universe'' – 390833306, of the Science and Technology Facilities Council (UK), grants ST/T006331/1 and ST/Y004515/1, as well as support by the German Volks\-wagen Stiftung and the European Research
Council (ERC) under the European Union’s Horizon 2020 research and innovation program Grant agreement No.\,948689.

\bibliographystyle{apsrev4-2} 
\bibliography{references.bib}

\end{document}